\documentclass[aps,prd,preprint,superscriptaddress,preprintnumbers,nobibnotes,showpacs]{revtex4}
\usepackage{amsfonts,amssymb,amsmath,bm}
\usepackage{graphicx}
\usepackage{pstricks,color}

\usepackage{graphicx}
\usepackage{dcolumn}
\usepackage{bm}
\def\br{\begin{eqnarray}}
\def\er{\end{eqnarray}}
\def\be{\begin{equation}}
\def\ee{\end{equation}}

\def\({\left(}
\def\){\right)}

\def\<{\left\langle}
\def\>{\right\rangle}

\begin{document}

\title{Vacuum energy as a c-function for theories  with \\ dynamically generated masses}

\author{A.~C.~Aguilar}
\affiliation{Federal University of ABC, CCNH, \\
Rua Santa Ad\'elia 166, 09210-170, Santo Andr\'e, Brazil\\}
\author{A.~Doff}
\affiliation{Universidade Tecnol\'ogica Federal do Paran\'a - UTFPR - COMAT \\
Via do Conhecimento Km 01, 85503-390, Pato Branco - PR, Brazil}
\author{A.~A.~Natale}
\affiliation{Instituto de F\'{\i}sica Te\'orica, UNESP - Universidade Estadual Paulista, \\
Rua Dr. Bento T. Ferraz, 271, Bloco II, 01140-070, S\~ao Paulo, Brazil}


\begin{abstract}
We argue that in asymptotically free non-Abelian gauge theories possessing the phenomenon of dynamical mass
generation the $\beta$ function is negative up to a value of the coupling constant that corresponds to a non-trivial fixed point, in
agreement with recent AdS/QCD analysis. This fixed
point happens at the minimum of the vacuum energy ($\Omega$), which, as a characteristic of theories with dynamical mass generation, 
has the properties of a c-function.
\end{abstract}


\maketitle

\par For unitary, renormalizable quantum field theories in two dimensions, Zamolodchikov has shown \cite{zam}
that there exists a positive-definite real function of the coupling constant $c(g)$ such that
\be
-\beta \frac{\partial}{\partial g} c(g,\mu) \leq 0 \,\, ,
\label{eq01}
\ee
where $\beta$ is the beta-function and $\mu$ is the renormalization scale. This means that there exists a
real function of the coupling constant that is monotonically decreasing along the renormalization group trajectories. 
The extension of the c-theorem to other dimensions was discussed by Cardy \cite{car}, where it was pointed all the difficulties to find 
such type of function in more than two dimensions. The attempts to demonstrate the existence of a c-theorem in four dimensions and its consequences led to several
studies (see, for example, \cite{app,ball,ken}), most of them following Cardy's proposal based on the Euler term in the trace of the
energy-momentum tensor and relating it to the conformal anomaly coefficient. 

It is particularly interesting the discussion of Ref.\cite{app}, where the renormalized free energy per unit volume, $\cal{F}$, 
is considered as a potential candidate for a c-function. In the high and low temperature limits the free energy can be used
to characterize the number of degrees of freedom, $f$, of the theory in the infrared and ultraviolet
regions: 
\be
f_{IR}\equiv - \lim_{T\rightarrow 0}\frac{{\cal{F}}}{T^4}\frac{90}{\pi^2} \,\,\,\,\, , \,\,\,\,\,
f_{UV}\equiv - \lim_{T\rightarrow \infty}\frac{{\cal{F}}}{T^4}\frac{90}{\pi^2} \,\, ,
\label{eq03}
\ee
where $T$ is the temperature. It was conjectured in Ref.\cite{app} that
\be
f_{IR} \leq f_{UV} \,\, ,
\label{eq02}
\ee
and this inequality was able to constrain the low energy structure of supersymmetric and non-supersymmetric
gauge theories \cite{app,app2}. A proof that the free energy, $\cal{F}$, plays the role of a c-function, as demonstrated
in Ref.\cite{app}, fails in the case when the low energy theory is a gauge field theory governed by a free infrared fixed
point. 

We will argue that the existence or not of a c-function in more than two dimensions may be a property related to the gauge
bosons and fermions dynamical mass generation (or breaking of the conformal symmetry), i.e. to theories where the mass generation 
mechanism is triggered by the non-trivial vacuum expectation value of composite operators. Therefore our discussion will make use
of the vacuum energy, $\Omega$, defined many years ago by Cornwall and Norton \cite{cn} and detailed in Ref.\cite{cjt}.
Of course, the vacuum energy and the free energy are not fully distinct quantities, but there is an extensive research on the $\Omega$
calculation as a function of the dynamical masses that will be quite useful for our purposes. We will show that $\Omega$ is a good
candidate for a c-function based on the fact that this quantity is always negative when a dynamical mass is generated, and we also
verify that in this case the $\beta$ function is negative, with the minimum of $\Omega$ happening at a non-trivial fixed point. 

The vacuum energy $\Omega = \Omega (g,\mu)$ can be defined as 
\begin{equation}
\Omega = V(G) - V_{pert}(G) \; ,
\label{e10}
\end{equation}
where we are subtracting from the effective potential for composite operators, $V(G)$, its perturbative counterpart,
and $V(G)$ is computed as a function of the nonperturbative Green functions (or complete propagators) $G$.
The effective potential is obtained from the effective action, $\Gamma (G)$, when we consider translationally invariant (t.i.) field
configurations, 
\[
V(G) \int d^4x = - \Gamma(G)|_{t.i.} \,\, . 
\]
A very important point is that $\Omega$ is
a finite function of its arguments, because the perturbative contribution has been subtracted out \cite{cn,cjt};
being a physical quantity its anomalous dimension vanishes and
$\Omega$ satisfies a simple homogeneous renormalization group equation~\cite{gross}
\begin{equation}
\left( \mu \frac{\partial}{\partial \mu} + \beta (g) \frac{\partial}{\partial g}
\right) \Omega = 0 \; .
\label{eq08}
\end{equation}

Consider gauge theories with dynamically generated gauge boson masses ($m$). In this case we can write
\mbox{$m =\mu f(g)$} \cite{gross}, from what follows that \mbox{$\mu (\partial m / \partial \mu ) = m$}
and, consequently,
\begin{equation}
m \frac{\partial \Omega}{\partial m} = - \beta(g)
\frac{\partial \Omega}{\partial g} \; .
\label{e3}
\end{equation}
However, as a physical parameter in $\epsilon$ dimensions, $\Omega$ has a simple scaling behavior
\be
\Omega (g,\mu) = \kappa m^\epsilon (g,\mu)\,\, ,
\label{eqx1}
\ee
where $\kappa$ is a calculable number, independent of $g$, consequently
\be
\epsilon \Omega = - \beta(g)
\frac{\partial \Omega}{\partial g} \; .
\label{eqx2}
\ee

In the sequence we shall discuss the following points: ({\textit{i}}) ${\partial \Omega}/{\partial m} =0$ is a
stationary point for the vacuum energy, meaning that the minimum of energy happens at a fixed point, because
${\partial \Omega}/{\partial g}\neq 0$ as demonstrated in Ref.\cite{we}; ({\textit{ii}}) $\Omega$ is negative
for theories with dynamically generated masses; therefore the right-hand side of Eq.(\ref{eqx2}) indicates that $\Omega$ is a c-function;
({\textit{iii}}) the $\beta$ function is always negative when we have dynamical mass generation. 

It is well known that the c-theorem holds in two dimensions;
therefore it would be interesting to check if the vacuum energy of a two dimensional theory, as the Gross-Neveu (GN) model,
satisfies the necessary requirements to be considered as a c-function. The GN model contains $N$ fermions and a four-fermion interaction with a coupling constant $g$, which can also be written in terms of an equivalent theory
with a scalar field $\sigma = g ({\bar{\Psi}}\Psi )$. In the leading order of the $1/N$ expansion, the effective
potential of this model is equal to
\be
V(\sigma )=\frac{1}{2}\sigma^2 + \frac{\lambda}{4\pi}\sigma^2\( ln\frac{\sigma^2}{\mu^2} -3 \) \,\, ,
\label{esig}
\ee
where $\lambda =g^2N$ is kept fixed, $\mu$ is a renormalization point, and the minima of $V$ occurs
at 
\[
|\sigma_m|=\mu exp(1-\pi/\lambda) \,\, . 
\]
The vacuum energy is simply 
\be
\Omega \equiv V(\sigma_m) \,\, ,
\ee
and the
chiral symmetry is dynamically broken generating a fermion mass 
\be
m\equiv m_F=g<\sigma>=\mu g\, exp(1-\pi/\lambda) \,\, .
\ee
As required by condition ({\textit{iii}}), the $\beta$ function is negative, more specifically \mbox{$\beta (g)=-(g^3N/2\pi )<0$}.
In addition, substituting the expression for the vacuum energy  in Eq.(\ref{e3}), we can  easily verify  that indeed $m(\partial \Omega/\partial m) < 0$
satisfying therefore the condition ({\textit{ii}}). Note that the fixed point structure of the GN model (when $\partial\Omega/\partial m\propto \beta(g) \rightarrow 0$) involves the knowledge of the $\beta$ function at higher orders in the $1/N$ expansion, when its vacuum energy depends on large $m_F$ values, but there are evidences for such critical coupling \cite{muta}.

\begin{figure}[!t]
\begin{center}
\includegraphics[scale=0.4]{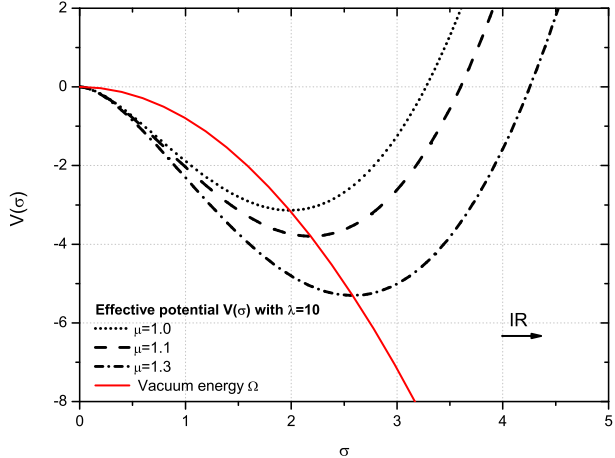}
\caption{Behavior of the vacuum energy in the case of the Gross-Neveu model. We show three curves of the effective potential
as a function of the field $\sigma$ and for different values of the renormalization point $\mu$. $\Omega$ is the curve
connecting all the minima and decreases towards the infrared values of the coupling constant.} 
\label{fig1}
\end{center}
\end{figure}

To make clear, the $\Omega$ behavior towards the infrared 
are illustrated in the Fig.~\ref{fig1}, where we show  that 
the expected behavior of the Gross-Neveu model effective potential as the coupling constant is changing
(increasing its value towards the IR). The minimum of energy, or $\Omega$ value, monotonically decreases as we go to larger and larger 
fermion masses (or larger values of the coupling constant).

Now, let us come back  for the case of non-abelian theory in four dimensions. Note that we will be dealing specifically with asymptotically 
free $SU(N)$ gauge theories where there is dynamical mass generation.
However this does not exclude the possibility to extend the vacuum energy as a c-function for other groups as well as to the case of
supersymmetric theories. In particular, all the SUSY cases treated in Ref.\cite{app} should follow the same idea discussed  here as long
as the theory develops different phases or vacuum expectation values.

It was demonstrated in Ref.\cite{we} that ${\partial \Omega}/{\partial g}\neq 0$ when the theory has a condensate or
develops a dynamically generated mass. Therefore Eq.(\ref{e3}) tell us that the $\beta$ function has a zero, i.e. a non-trivial fixed point, at the
minimum of the vacuum energy (${\partial \Omega}/{\partial m} =0$). This point corresponds to the equality of
Eq.(\ref{eq01}) when $\Omega$ plays the role of a c-function. Away from the fixed point the inequality ($- \beta(g)
[{\partial \Omega}/{\partial g}]<0 $) is due to the fact that a non-trivial dynamics leading to 
mass generation lowers the vacuum energy, i.e. $\Omega$ \textsl{is negative}, as we shall discuss in detail ahead, 
in such a way that the left-hand side of Eq.(\ref{eqx2}) is
negative, therefore $ - \beta(g)[{\partial \Omega}/{\partial g}]\leq 0$ and $\Omega$ is a c-function. The inversion method 
discussed in Ref.\cite{we} can also be used to show that $\partial \Omega /\partial g$ is monotonic away from the fixed point.

To show that the $\beta$ function is negative in the case of non-Abelian gauge theories with 
dynamical gauge boson mass generation, we may particularize the problem to the pure gauge theory.
In this case the Lagrangian is given by ${\cal L}= \frac{1}{2}G_{\mu\nu}^2$. We can now rescale the fields $A_\mu^a
=g^{-1}{\hat{A}}_\mu^a$, $G_{\mu\nu}^a =g^{-1}
{\hat{G}}_{\mu\nu}^a$, and regularize the vacuum energy (and the
potential) setting its perturbative part equal to zero in order
to obtain the generating functional ($Z$) \cite{cornwall}
\begin{eqnarray}
Z &=& Z_p^{-1} \int d{\hat{A}}_\mu
\exp \left[ -g^{-2} \int d^4 x\frac{1}{4}  \sum_a ({\hat{G}}_{\mu\nu}^a)^2
\right]\nonumber \\
&=& e^{-V \Omega} \; ,
\label{e11}
\end{eqnarray}
where $V$ is the volume of Euclidean space-time and $Z_p$ is the perturbative functional.
Differentiating with respect to $g$ it follows that
\begin{equation}
\frac{\partial \ln Z}{\partial g} = \frac{1}{2g} \int d^4 x
\< \sum_a ({\hat{G}}_{\mu\nu}^a)^2 \>_{reg} = - \frac{V\partial\Omega}{\partial g},
\label{e12}
\end{equation}
where $\left\langle  \sum_a ({\hat{G}}_{\mu\nu}^a)^2\right\rangle$ is the gauge boson condensate and the subscript $reg$ indicates that the regularization is by
subtraction of the perturbative expectation value in the same way
as indicated in Eq.(\ref{e10}). The factor $V$ in the right-hand
side is canceled with the one coming out from the $x$
integration. It must be noted that Eq.(\ref{e12}) can be related to the Euclidean trace anomaly \mbox{($<\theta> = -4\Omega$)} and
that $-(\partial\Omega/\partial g)$ \textsl{is a positive quantity} \cite{app,cornwall}. Therefore, looking
at Eq.(\ref{eqx2}), necessarily the $\beta$ function is negative, as affirmed in the item ({\textit{iii}}) above.

A central point in our discussion is the fact that $\Omega$ is negative, or, as we discussed before,
the presence of dynamically generated masses lowers the vacuum energy. It is well known that in asymptotically 
free non-Abelian gauge theories the vacuum has a minimum of energy  when dynamical fermion masses are generated, which is
supported, in the QCD case, by the chiral symmetry symmetry breaking phenomenology. Less known is the fact that
such theories also generate gauge boson dynamical masses. It has already been
observed through a variational calculation in four dimensional $SU(N)$ theories,
that the vacuum energy is best minimized by a variational state characterized by a dynamically
generated mass scale $m$ \cite{kogan}, indicating that $\Omega$ is indeed negative. 
In a different approach we can minimize the effective potential for
composite operators (and consequently the vacuum energy) up to two loops, obtaining the
Schwinger-Dyson equations for the gauge boson propagators. These equations can be reorganized in a gauge
invariant (transversal) formulation, truncated and solved under certain approximations, resulting in a  
dynamically massive solution \cite{cornwall,papa}, which is in agreement with 
lattice simulations of $SU(2)$ and $SU(3)$ gauge theories in three and four dimensions \cite{aguilar}.

A dynamical gauge boson mass displays the following asymptotic behavior
\[
m(q^2) \rightarrow m \,\, (\equiv const) \,\,\,\,\,  as \,\,\,\,\, q\rightarrow +0 \,\, ,
\]
and
\[
m(q^2) \rightarrow 1/q^\eta \,\,\,\,\, as  \,\,\,\,\, q\rightarrow +\infty \,\, ,
\]
where $\eta$ is some constant calculable from the SDE solution (obtained from the condition $\partial \Omega / \partial m = 0$). For QCD it has been shown that $\eta =2$ \cite{lavelle,aguilar,papa}. The massive 
solution obtained in this procedure indeed minimizes the vacuum
energy in the Hartree approximation \cite{cornwall,gorbar}. This 
decreasing behavior with the momentum is typical of any dynamically generated mass and preserves
unitarity. 

The presence of the dynamically generated mass also modifies the IR behavior of the
QCD running coupling, ${\alpha}_{s}(q^2)$, that can be modelled as 
\be 
{\alpha}_{s}(q^2) = \frac{1}{4\pi b \ln [(q^2 + 4m^2(q^2))/\Lambda_{\rm QCD}^2]} \,\, ,
\label{e1a}
\ee 
where $b  = 11C_{\rm A}/48\pi^2$
is the first coeficient of the $\beta$-function, $C_{\rm {A}}$ is the Casimir eigenvalue of the adjoint representation
($C_{\rm {A}}=N$ for $SU(N)$), and $\Lambda_{\rm QCD}$ is the characteristic QCD mass scale of a few hundred ${\rm MeV}$.

\begin{figure*}[!t]
\begin{minipage}[b]{0.45\linewidth}
\centering
\includegraphics[scale=0.4]{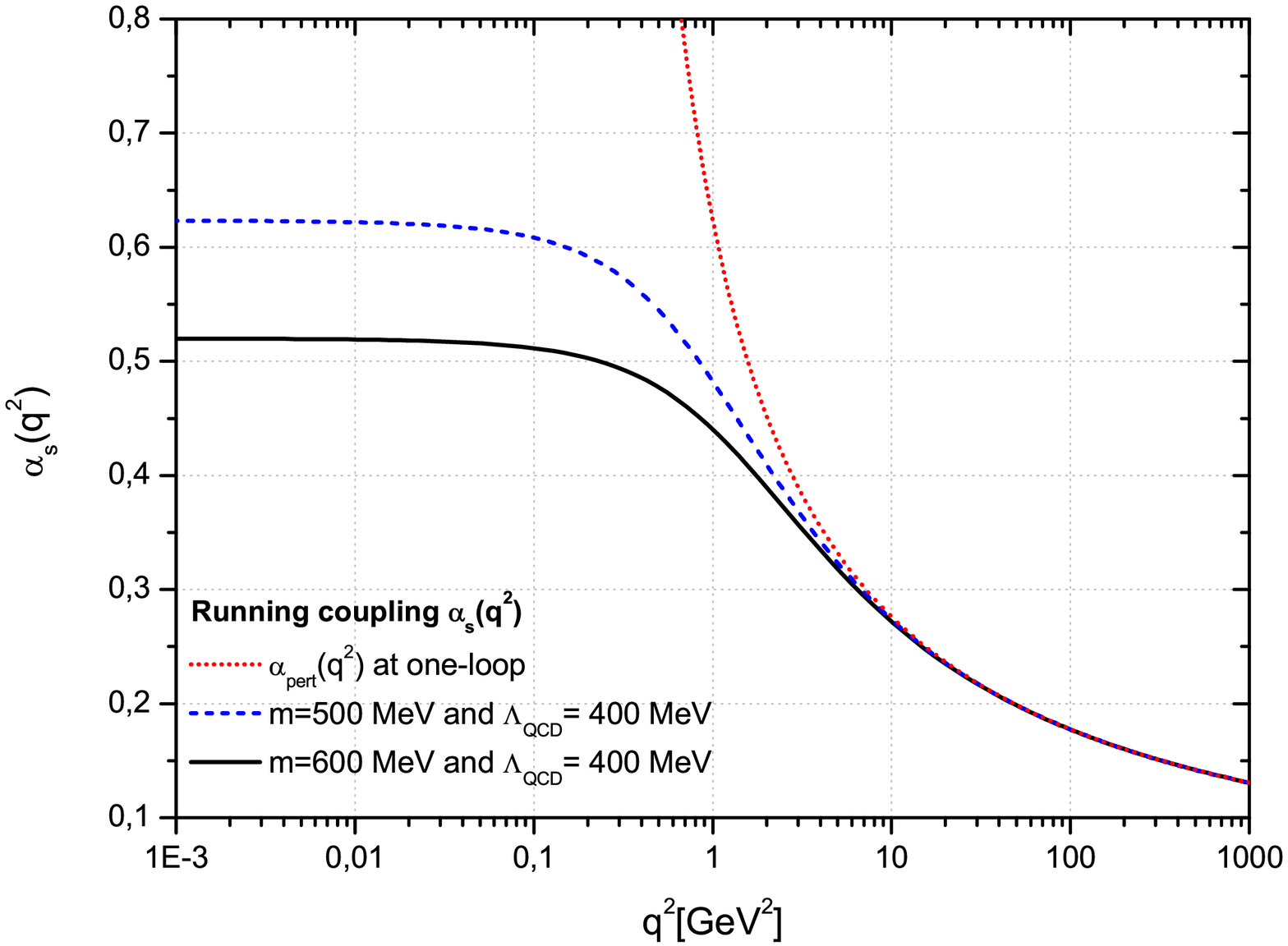}
\end{minipage}
\hspace{0.5cm}
\begin{minipage}[b]{0.45\linewidth}
\centering
\includegraphics[scale=0.4]{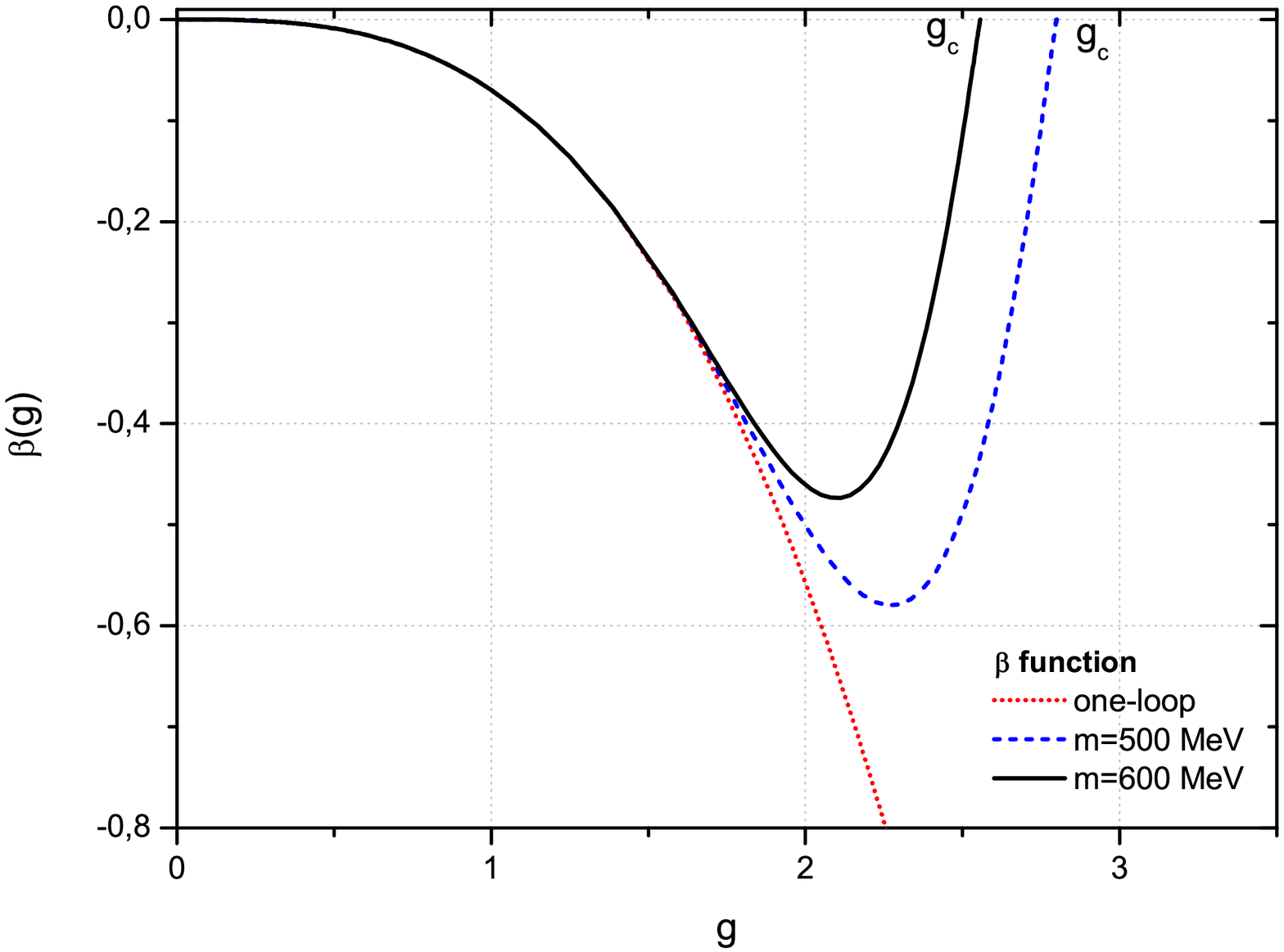}
\end{minipage}
\caption{{\it Left panel}:  The IR finite QCD running coupling, $\alpha_s(q^2)$, given by Eq.(\ref{e1a}) 
for a  gluon  mass of \mbox{$m=500$ MeV} (dashed blue curve) and \mbox{$m=600$ MeV}(continuous black curve)
when \mbox{$\Lambda_{\rm QCD} = 400$ MeV}. The dotted red curve represents the one-loop perturbative behavior $\alpha_{pert}(q^2)$. 
{\it Right panel}: The corresponding $\beta$ function, defined as \mbox{$\beta(g) = q (dg(q)/dq)$}, when \mbox{$m=500$ MeV} (dashed blue curve)  
and \mbox{$m=600$ MeV} (continuous black curve) compared to the one-loop perturbative value, $-bg^3$ (dotted red curve).}
\label{fig2}
\end{figure*}

From Eq.(\ref{e1a}), we can easily see that  ${\alpha}_{s}(q^2)$ saturates 
in the deep infrared, reaching a finite value at $q^2=0$, which depends only on 
the ratio $m/\Lambda_{\rm QCD}$ \cite{cornwall,cornwall2}. 
In the left panel of Fig.~\ref{fig2}, we show ${\alpha}_{s}(q^2)$ for two different
values of $m$, where we see that the higher is the ratio $m/\Lambda_{\rm QCD}$ the lower will be the value  of
${\alpha}_{s}(0) =g_c^2/4\pi$. Moreover,
for asymptotically large 
momentum, we recover  the one-loop perturbative behavior $\alpha_{pert}(q^2)$, represented 
by the dotted red curve.


Notice that the appearance of $m^2(q^2)$ not only allows for 
an infrared finite value for  ${\alpha}_{s}(q^2)$ but it also tames  the Landau pole.
As a consequence, such class of theories are not plagued by renormalon ambiguities, and calculations as the
one defined by Eq.(\ref{e10}), where the finite result comes out by
subtraction of the perturbative contributions are not ill-defined. Such theories
may have a skeleton expansion, where the
freezing of the running coupling constant at low energy scales could allow to
capture at an inclusive level the nonperturbative effects in a reliable way as argued in Ref.\cite{brod}
(without renormalon ambiguities). The first steps to obtain such expansion 
are outlined in the last work in Ref.\cite{papa}.

We now discuss a specific calculation of the vacuum energy for pure gauge theories.
Cornwall \cite{cornwall} was the first one to compute $\Omega$ in the QCD case with dynamically massive gluons, and
these calculations are expected to be valid for any non-Abelian asymptotically 
free gauge theory. The vacuum energy, $\Omega$, was calculated in the
Hartree approximation for pure $SU(N)$ gauge theories, assuming hard gauge boson masses (i.e. neglecting all momentum dependence in the masses), with the divergent integral regulated as described in Ref.\cite{cornwall}. Note that these divergences appear only because the running of the
masses is not considered (otherwise $\Omega$ is finite), the result is 
\br
\Omega (m,g)=
-\frac{3(N^2-1)}{2(2\pi)^4} \int d^4k\,\, \left[\ln\left(\frac{k^2+m^2}{k^2}\right) -\frac{m^2}{k^2+m^2} \right] 
-\frac{3(N^2-1)}{4Ng^2}m^4 \,.
\label{eq202}
\er
It is easy to observe that $\Omega$ is negative, which
again confirms the assertion of item ({\textit{ii}}) above. Therefore in theories with dynamically generated gauge boson masses $\Omega$ is negative,
${\partial \Omega}/{\partial m}=0$ indicates the presence of a fixed point at $g_c$ \cite{we}, the $\beta$ function is negative and near
the fixed point the $\beta$ function may behave as $\beta \propto (g_c - g)^{1/2}$ \cite{cornwall2} (See the right
 panel of Fig.~\ref{fig2}). Notice that the same qualitative behavior for the $\beta$ function is found in Ref.\cite{brodsky}.

The introduction of fermions in the non-Abelian gauge theory does not modify our arguments. We can follow Ref.\cite{gorbar}
to compute the fermionic vacuum energy, adding to $\Omega$ the fermionic contribution ($\equiv\Omega_f$),
which, in one approximation where we neglected the running of dynamical fermion masses ($m_f$), is equal to
\br
\Omega_f \approx && 2Nn_f 
\int d^4k \left[-\ln\left(\frac{k^2+m_f^2}{k^2}\right)
+\frac{m_f^2}{k^2+m_f^2}+\frac{m_f^4}{2k^4} \right] \; .
\label{eq21}
\er
The calculation was performed for a $SU(N)$ gauge theory with $n_f$ fermions. 
We are assuming that the dynamical gauge boson mass ($m$) is larger than the dynamical fermion mass ($m_f$). This means that the fermionic contribution is just a small perturbation in the full vacuum, and allows the simple sum of different contributions to be reliable despite the
rough approximations to compute the effective action \cite{gorbar}. Of course, this is justified as long as we maintain asymptotic freedom, in such a way that the instability due to the fermionic contribution is small, and also by the fact that, at least in QCD, the dynamical gauge boson masses are twice the fermionic ones (see, for instance, Ref.\cite{cornwall,nat} where it is pointed out that $m\approx 2\Lambda_{\rm QCD}$ whereas we expect $m_f\approx \Lambda_{\rm QCD}$). 

It can be verified that the vacuum energy that we discussed up to now, when computed at finite temperature, is equivalent to the renormalized
free energy discussed in Ref.\cite{app}, as it is possible to see if we compare the free energy obtained by Freedman and McLerran \cite{fmc} with the loop expansion of Ref.\cite{cjt,cornwall3}. We expect that at very high temperatures all dynamical masses are erased and we end up with an
almost non-interacting $SU(N)$ theory with $(N^2-1)$ gauge bosons and $n_f$ fermions. The vacuum or free energy will just be $kT$ factors
times the number of degrees of freedom, which will be $[2(N^2-1)+\frac{7}{8}(4Nn_f)]$. 

At low temperatures we shall not have exactly
the vacuum energy that we calculated above, because confinement is missing from the Green's functions in the above calculation. However
we know that the vacuum energy will be approximated by the same $kT$ factors times $[(n_f^2-1)]$, which is the number of degrees of freedom of the Goldstone bosons resulting from the chiral ($SU(n_f)\times SU(n_f)$) symmetry breaking. This is the dominant contribution to the vacuum energy, but it must be remembered that all the other excitations (fermionic as well as the ones formed by gauge bosons) also contribute to the vacuum energy at higher order. In Ref.\cite{app} it was obtained a constraint
on the number of fermions, that for QCD reads, $n_f < 12$; therefore, taking into account the effects of other excitations that we
discussed above, which, unfortunately, have to be computed with the help of models for low temperature QCD, since we do not know how to
introduce confinement in the $\Omega$ calculation, may lead to an even tighter constraint on the number of fermions. Notice that the vacuum energy also takes into account
the thinning of degrees of freedom as we move towards the infrared values of the coupling constant, similarly to what is
usually discussed in the standard renormalization group framework.

We may have other constraints on the particle spectrum without the need of comparing the extreme $T\rightarrow 0$ and
$T\rightarrow\infty $ limits. The confinement and chiral symmetry breaking phase transitions separate different regions of the
vacuum energy for asymptotically free non-Abelian gauge theories. In the QCD case it is expected that the confinement and chiral transitions for quarks in the fundamental representation
are quite close, while the chiral transition for quarks in the adjoint representation would happens at a temperature different
of the confinement one \cite{cornwall4}. As the coupling constant increases with the decrease of the temperature we may have a specific phase transition
happening at some critical temperature ($T_c$), and we may find temperature regions above ($T_a >T_c$) and below  ($T_b <T_c$) the critical
temperature such that 
\be
\Omega_{T_b} \leq \Omega_{T_a} \,\, .
\ee

At zero temperature we can also expect that gauge theories with similar $\beta$ function, i.e. with approximately the same value for the
coefficients of the $\beta$ function for the same gauge group, but fermions in different representations, for instance in the
fundamental ($f$) and adjoint representation ($A$), will have \cite{doff}
\be
\Omega^{(f)} \leq \Omega^{(A)} \,\,. 
\ee

This is a consequence of the
monotonic behavior of $\Omega$ with the coupling constant and the fact that the vacuum energy
(or the effective potential for composite operators that generates the gap equation) scales with the coupling constant
times the Casimir operator ($\propto g^2 C$). 

In conclusion, we are arguing that asymptotically free non-Abelian gauge theories possessing the phenomenon of
dynamical mass generation, for gauge bosons and fermions, have a fixed point at the minimum of the vacuum energy, the
$\beta$ function is negative up to the fixed point and the vacuum energy is a good candidate for a c-function. The $\beta$ function
behavior for theories with dynamically generated gauge boson and fermion
masses displays the same qualitative behavior as the one found in an AdS/QCD analysis \cite{brodsky}.
We believe that the c-theorem can be better understood as a property of theories with dynamical mass generation, and, consequently, be related
in this way to the breaking of the conformal symmetry.

\acknowledgments

We are grateful to J. Papavassiliou for the critical reading of the manuscript. This research was partially supported by the Conselho Nacional de Desenvolvimento
Cient\'{\i}fico e Tecnol\'ogico (CNPq).

\end{document}